\documentclass[prc,aps,eqsecnum,floatfix,showpacs]{revtex4}
\usepackage{dcolumn}
\usepackage{bm}
\usepackage{graphicx}
\begin{document}
\title{Neutron spin rotation in $\vec{n}$-$d$ scattering}
\author{R.\ Schiavilla$^{1,2}$, M.\ Viviani$^3$, L.\ Girlanda$^3$,  A.\
  Kievsky$^3$, L.E.\ Marcucci$^3$ } 
\affiliation{
$^1$\mbox{Jefferson Lab, Newport News, VA 23606, USA} \\
$^2$\mbox{Department of Physics, Old Dominion University, Norfolk, VA 23529, USA} \\
$^3$\mbox{INFN, Sezione di Pisa, and Department of Physics, University of
Pisa, I-56127 Pisa, Italy}
}
\date{\today}
\begin{abstract}
The neutron spin rotation induced by parity-violating (PV) components in the
nucleon-nucleon potential is studied in $\vec{n}$-$d$ scattering at zero
energy.  Results are obtained corresponding to the Argonne $v_{18}$
two-nucleon and Urbana-IX three-nucleon strong-interaction potentials in combination
with either the DDH or pionless EFT model for the weak-interaction potential.
We find that this observable is dominated by the contribution of the long-range
part of the PV potential associated with pion exchange.  Thus its measurement could provide a
further constraint, complementary to that coming from measurements of the photon
asymmetry in $\vec{n}$-$p$ radiative capture, on the strength of this component
of the hadronic weak interaction.
\end{abstract}
\maketitle
\section{Introduction}
\label{sec:intro}

We report on a calculation of the neutron spin rotation in $\vec{n}$-$d$
scattering at zero energy.  The present work is the
continuation of a program to study parity-violating (PV) effects in few-nucleon
systems induced by hadronic weak interactions.  Two earlier papers dealt with the
two-nucleon systems: the first~\cite{Carlson02} was devoted to $\vec{p}$-$p$ elastic
scattering, and presented a calculation of the longitudinal asymmetry in the
lab-energy range 0--350 MeV.  The second~\cite{Schiavilla04} provided a detailed
account of the i) neutron spin rotation in $\vec{n}$-$p$ scattering at zero
energy, ii) longitudinal asymmetry in $\vec{n}$-$p$ elastic scattering up to lab
energies of 350 MeV, iii) photon angular asymmetry in $\vec{n}$-$p$ radiative
capture at thermal neutron energies, and iv) photon helicity dependence of
the $d(\vec{\gamma},n)p$ cross section from threshold up to energies of 20 MeV.

In this study we adopt two different models for the PV potential.  One is
parameterized in terms of $\pi$-, $\rho$-, and $\omega$-meson exchanges, in
which the weak couplings of these mesons to the nucleon are estimated
within a quark-model approach incorporating symmetry arguments and current
algebra requirements.  It is known as the DDH model~\cite{Desplanques80},
and has provided for over two decades a useful framework for the analysis
and interpretation of measurements of PV observables (see in this connection
the recent review in Ref.~\cite{Ramsey06}).

The other model is that derived recently by Zhu {\it et al.}~\cite{Zhu05} in an
effective field theory (EFT) approach, in which only nucleons are retained as
explicit dynamical degrees of freedom, while pions and baryon resonances, such as
$\Delta$ isobars, are integrated out.  It is formulated in terms of a number of
four-nucleon contact terms.  Obviously, its use
is restricted to processes occurring at energies much below the pion mass.

The rotation of the neutron spin in a plane transverse to the beam
direction is calculated perturbatively to first order in the DDH and
EFT PV potentials.  We use $n$-$d$ wave functions obtained
with the hyperspherical-harmonics (HH) method~\cite{Kievsky94,Kievsky04,Kievsky08} from
realistic strong-interaction Hamiltonians, consisting of the Argonne
$v_{18}$~\cite{Wiringa95} (AV18) two-nucleon potential with and without
the inclusion of a three-nucleon potential, the Urbana IX model~\cite{Pudliner95}.
The AV18/UIX Hamiltonian quantitatively and successfully accounts for a
wide variety of few-nucleon bound-state properties and reactions, ranging
from binding energies and charge radii to electromagnetic form factors,
low-energy scattering observables and electroweak capture cross sections
(for a review, see Refs.~\cite{Carlson98,Marcucci06}).

Our prime objectives in this as in the earlier papers are to
develop a systematic and consistent framework for studying PV 
observables in few-nucleon systems, where accurate microscopic
calculations are feasible, and to use available and
forthcoming experimental data on these observables to constrain
the strengths of the long- and short-range parts of the two-nucleon
weak interaction.  While no measurements of the neutron spin
rotation in $\vec{n}$-$d$ scattering have been carried out so far
to the best of our knowledge, nevertheless we hope the present
study will provide strong motivation for undertaking this type
of experiments.  Within the context of the calculations based on
the DDH model, we find that this observable is dominated by the
contribution associated with the long-range pion-exchange component.
Indeed, assuming a liquid deuterium density of $\rho=0.4\times10^{23}$ atoms
cm${}^{-3}$ and the ``best'' values of Ref.~\cite{Desplanques80}
(as listed in Table II of Ref.~\cite{Schiavilla04}) for the $\pi$-,
$\rho$-, and $\omega$-meson weak (and strong) coupling constants to
the nucleon, the predicted spin rotation per unit-length of matter
traversed is $ 0.52\times 10^{-7} $ rad-cm$^{-1}$ 
with the complete DDH potential, and
$ 0.46 \times 10^{-7}$ rad-cm$^{-1}$ when only its pion-exchange term is retained.
We find the model dependence of this result upon the (realistic)
strong-interaction potentials used to generate the $n$-$d$ continuum
states to be quite small.  Thus, a measurement of the $\vec{n}$-$d$ spin
rotation could provide a further constraint, complementary to that
coming from the measurements in progress of the photon asymmetry in
the $\vec{n}$-$p$ radiative capture, on the strength of the long-range
part of the PV potential.

The remainder of this paper is organized as follows.  In
Sec.~\ref{sec:nrot} the neutron-spin-rotation observable is
related to matrix elements of the PV potential between incoming
and outgoing $n$-$d$ states.  The DDH
and EFT PV potential models are briefly discussed in Sec.~\ref{sec:pv}; in particular,
in Appendix~\ref{app:fierz} it is shown that the number of independent
four-nucleon contact terms in the EFT formulation is five rather than
ten, as obtained in Ref.~\cite{Zhu05}.  The Monte Carlo techniques
used to calculate the relevant matrix elements are reviewed in Sec.~\ref{sec:cal},
where the details of a test calculation based on simple Gaussian
wave functions are also described.  Finally, Sec.~\ref{sec:res} contains
a brief discussion of the results along with some concluding remarks.

\section{The neutron spin rotation observable}
\label{sec:nrot}
In first order perturbation theory in the weak interactions,
the neutron spin rotation per unit-length of matter traversed,
${\rm d}\phi/{\rm d}z$, in $\vec n$-$d$ scattering is given
by~\cite{Schiavilla04,Fermi50}

\begin{equation}
\frac{1}{\rho}\frac{{\rm d}\phi}{{\rm d}z} = \frac{1}{3\, v_{\rm rel}} {\rm Re} 
\! \sum_{m_n m_d} \epsilon_{m_n}\, ^{(-)}\langle p\hat{\bf z}; m_n , m_d \mid v^{\rm PV} \mid 
p\hat{\bf z};m_n , m_d\rangle^{(+)}\ ,
\label{eq:nphi}
\end{equation}
where $\rho$ is the density of deuterons (a parameter under control
of the experimenter), $v^{\rm PV}$ denotes the PV nuclear
potential, $\mid\! p \hat{\bf z};m_n,m_d\rangle^{(\mp)}$
are the $n$-$d$ scattering states with incoming-wave $(-)$ and
outgoing-wave $(+)$ boundary conditions and relative momentum
${\bf p}=p \hat{\bf z}$ taken along the spin-quantization axis,
{\it i.e.} the $\hat{\bf z}$-axis, and $v_{\rm rel}=p/\mu$ is
the magnitude of the relative velocity, $\mu$ being the $n$-$d$
reduced mass.  The expression above is averaged over the spin
projections $m_d=\pm 1, 0$ of the deuteron, however, the phase
factor $\epsilon_{m_n}\equiv (-)^{1/2-m_n}$ is $\pm 1$ depending
on whether the neutron has $m_n=\pm 1/2$. 

The scattering states are calculated from realistic (strong
interaction) Hamiltonians including two- and three-nucleon
potentials by means of accurate hyperspherical-harmonics
techniques~\cite{Kievsky94,Kievsky04,Kievsky08}, and are expanded in partial
waves as~\cite{Viviani96} 
\begin{equation}
\mid p\hat{\bf z};m_n m_d\rangle^{(\pm)}=
\sqrt{4\pi}\! \sum_{L S J} i^L\, \sqrt{2L+1}\, 
\langle 1/2 \, m_n,1\, m_d\mid SJ_z\rangle\,
\langle SJ_z,L\, 0\mid JJ_z\rangle\, \mid p;LS;J,J_z \rangle^{(\pm)} \ ,
\end{equation}
where
\begin{equation}\label{eq:psipm}
  \mid p;LS;J,J_z\, \rangle^{(\pm)} = \sum_{L^\prime S^\prime} 
  \left[ 1\mp {\rm i}\, ^J\!R(p)\right]_{LS,L^\prime S^\prime}^{-1}  \, \mid \overline{p;L^\prime
  S^\prime; J,J_z}\rangle \ .
\end{equation}
In these equations, $L$ denotes the relative orbital angular momentum
between the two clusters, $S$ their channel spin (either 1/2 or 3/2),
and $[^J\!R(p)]_{LS,L^\prime S^\prime}$ is the (real) $R$-matrix
at center-of-mass energy $p^2/(2\mu)$, from which phase-shifts and
mixing angles in the channel specified by total angular momentum $J$
are easily obtained.  Lastly, the state $\mid \overline{p;LS; J,J_z}\rangle$
is defined as in Eqs.~(2.8) and (2.9) of Ref.~\cite{Viviani96}, and in
the asymptotic region (at large $n$-$d$ separations) reduces to
\begin{eqnarray}
 \mid \overline{p;LS; J,J_z}\rangle &\rightarrow& \frac{1}{\sqrt{3}}
 \sum_{ijk\, {\rm cyclic}} \sum_{L^\prime S^\prime} w^J_{LS,L^{\prime}S^{\prime}}(p;y_i)
 \Big[ Y_{L^\prime}(\hat{\bf y}_i)
\, \otimes \mid i,jk; S^\prime \rangle \Big]_J  \ ,
\label{eq:psia} \\
w^J_{LS,L^{\prime}S^{\prime}}(p;y_i) &=& \delta_{LL^\prime}\,
 \delta_{SS^\prime}\, j_{L^\prime}(py_i) +\, ^J\!R_{LS,L^\prime S^\prime}(p)\,
 n_{L^\prime}(py_i) \ ,
\label{eq:psiab}
\end{eqnarray}
where $\mid i,jk; S^\prime \rangle$ denotes a state in which neutron $i$
and a deuteron made up of nucleons $jk$ are coupled to channel spin $S^\prime$,
$j_L(x)$ and $n_L(x)$ are, respectively, the regular and irregular
spherical Bessel functions, and ${\bf y}_i={\bf r}_i-({\bf r}_j+{\bf r}_k)/2$
is the $n$-$d$ vector separation.

We are interested in very low-energy neutrons, and thus
it is sufficient to keep only S- and P-wave channels in
the partial wave expansion of $\mid p\hat{\bf z};m_n,m_d\rangle^{(\pm)}$.
Equation~(\ref{eq:nphi}) is then simply written as
\begin{eqnarray}
 \frac{1}{\rho}\frac{{\rm d}\phi}{{\rm d}z} &=&\frac{8 \pi}{\sqrt{3} v_{\rm rel}}
 \sum_{m_n m_d}\sum_{SJ}\epsilon_{m_n}\, \langle 1/2 \, m_n,1\, m_d\mid JJ_z\rangle\,
\langle 1/2 \, m_n,1\, m_d\mid SJ_z\rangle \, \nonumber \\
&&\langle SJ_z,1\, 0\mid JJ_z\rangle \,
{\rm Im}\Big[ {^{(-)}\!\langle} \,p;1S;J,J_z\,\mid v^{\rm PV} 
\mid p;0J; J,J_z\,\rangle^{(+)}\Big] \ .
\label{eq:eq1}
\end{eqnarray}
The relation above follows by first noting that under time
inversions, induced by the antiunitary operator ${\cal T}$,
the states in the convention adopted here transform as
\begin{equation}
{\cal T}\, \mid p;LS; J,J_z\,\rangle^{(+)} = (-)^{L+J-J_z}\,
\mid p;LS; J,-J_z\,\rangle^{(-)} \ ,
\end{equation}
and then using the fact that the PV potential is i) a scalar under
rotations (and hence its matrix elements are independent of $J_z$)
and ii) invariant under time inversions, which lead to
\begin{equation}
{^{(-)}\!\langle} \, p;0J;J,J_z\,\mid v^{\rm PV} 
\mid p;1S; J,J_z\,\rangle^{(+)} =- \,\,
{^{(-)}\!\langle} \, p;1S;J,J_z\,\mid v^{\rm PV} 
\mid p;0J; J,J_z\,\rangle^{(+)}  \ .
\end{equation}

In Eq.~(\ref{eq:eq1}) the sums over $S,J$ are restricted to $S,J=1/2$ and 3/2,
and thus the PV potential connects the doublet (quartet) S-wave state
$^2$S$_{1/2}$ ($^4$S$_{3/2}$) to both the doublet and quartet P-wave
states $^2$P$_{1/2}$ ($^2$P$_{3/2}$) and $^4$P$_{1/2}$ ($^4$P$_{3/2}$).
Note that the additional contribution associated with the transition
$^4$S$_{3/2} \rightarrow\, ^4$F$_{3/2}$ is neglected at the low energies
of interest here.
\section{The parity-violating potential}
\label{sec:pv}

Two different models of the PV weak-interaction potentials are
adopted in the calculations reported below.  One is the model
developed almost thirty years ago by Desplanques
{\it et al.}~\cite{Desplanques80} (and known as DDH): it
is parameterized in terms of $\pi$-, $\rho$-, and $\omega$-meson
exchanges, and involves in principle seven weak pion and
vector-meson coupling constants to the nucleon.  These
were estimated within a quark model approach incorporating
symmetry arguments and current algebra
requirements~\cite{Desplanques80,Holstein81}.  Due to
the inherent limitations of such an analysis, however,
the coupling constants determined in this way have rather wide
ranges of allowed values.

The other model for the PV potential considered in the present
work is that formulated in 2005 by Zhu {\it et al.}~\cite{Zhu05}
within an effective-field-theory (EFT) approach in which
only nucleon degrees of freedom are retained explicitly.
At lowest order $Q/\Lambda_\chi$, where $Q$ is the small
momentum entering the low-energy PV process and $\Lambda_\chi
\simeq 1$ GeV is the scale of chiral symmetry breaking,
it is parameterized by a set of twelve contact four-nucleon terms.
In fact, it has been recently realized~\cite{Girlanda08}
that this set is redundant, and the number of independent
contact terms can be reduced to five by Fierz-rearrangement.  The
derivation is summarized in Appendix~\ref{app:fierz}.

The DDH and EFT PV two-nucleon potentials are conveniently written as
\begin{equation}
 v^\alpha_{ij}=\sum_{n=1}^{13} c^\alpha_n \, O^{(n)}_{ij} \ ,
  \qquad \alpha={\rm DDH} \,\, {\rm or}\,\, {\rm EFT} \ ,
\end{equation}
where the parameters $c^\alpha_n$ and operators $O^{(n)}_{ij}$, $n=1,\dots, 13$,
are listed in Table~\ref{tb:tab1}.  In this table the vector operators
${\bf X}^{(n)}_{ij,\pm}$ are defined as
\begin{eqnarray}
 {\bf X}^{(n)}_{ij,+} &\equiv& \left[ {\bf p}_{ij} \, ,\, f_n(r_{ij}) \right]_+ \ , 
\label{eq:x+} \\
 {\bf X}^{(n)}_{ij,-} &\equiv& {\rm i}\, \left[ {\bf p}_{ij} \, ,\, f_n(r_{ij}) \right]_- \ ,
\label{eq:x-}
\end{eqnarray}
where $\left[ \dots \, , \, \dots \right]_{\mp}$ denotes the commutator ($-$)
or anticommutator ($+$), and ${\bf p}_{ij}$ is the relative momentum operator, 
${\bf p}_{ij} \equiv ({\bf p}_i-{\bf p}_j)/2$.  In the DDH model,
the functions $f_x(r)$, $x=\pi, \rho$ and $\omega$, are Yukawa functions,
suitably modified by the inclusion of monopole form factors,
\begin{equation}
 f_x(r) =\frac{1}{4\pi\, r} \left\{
 {\rm e}^{-m_x r} -{\rm e}^{-\Lambda_x r}\left[1+
 \frac{\Lambda_x r }{2}\left(1-\frac{m_x^2}
 {\Lambda_x^2}\right)\right] \right\} \ .
\label{eq:fx}
\end{equation}
In the EFT model, however, the short-distance behavior is
described by a single function $f_\mu(r)$, which is
itself taken as a Yukawa function with mass parameter $\mu$,
\begin{equation}
 f_\mu(r)=\frac{1}{4\pi\, r}{\rm e}^{-\mu r} \ ,
\label{eq:fmu}
\end{equation}
with $\mu \simeq m_\pi$ as appropriate in
the present formulation, in which pion degrees of
freedom are integrated out.

In $v^{\rm DDH}_{ij}$, the strong-interaction coupling constants
of the $\pi$-, $\rho$-, and $\omega$-meson to the nucleon are denoted
as $g_\pi,\, g_\rho,\, \kappa_\rho,\, g_\omega,\, \kappa_\omega$,
while the weak-interaction ones as $h^1_\pi,\, h^0_\rho,\, h^1_\rho,\,
h^{\prime 1}_\rho,\, h^2_\rho,\, h^0_\omega,\, h^1_\omega$, where
the superscripts 0, 1, and 2 specify the isoscalar, isovector,
and isotensor content of the corresponding interactions.
In the EFT model, the five low-energy constants $C_1, \tilde{C}_1,
C_2+C_4 , C_5$ and $C_6$
completely characterize $v^{\rm EFT}_{ij}$, to lowest order
$Q/\Lambda_\chi$ (see Appendix~\ref{app:fierz}).

In the analysis of the neutron spin-rotation observable to follow,
we will report results for the coefficients $I^{\rm DDH}_n$ and
$I^{\rm EFT}_n$ in the expansion
\begin{equation}
\frac{1}{\rho}\frac{{\rm d}\phi}{{\rm d}z} =
\sum_{n=1}^{13} c^\alpha_n \, I^\alpha_n \ . 
\label{eq:in}
\end{equation}
Thus we will not need to consider specific values, or rather
range of values, for the strength parameters $c^\alpha_n$.
However, the $I^\alpha_n$ depend on the masses (and short-range
cutoffs $\Lambda_x$ for the DDH model) occurring in the Yukawa
functions.  Two different sets of values are used for $\Lambda_x$
and $\mu$ in the present study, as listed in Table~\ref{tb:tab2}.
Both were used in the DDH-based calculations of PV two-nucleon
observables in Refs.~\cite{Carlson02,Schiavilla04}.
\begin{table}[t]
\begin{tabular}{c|c|c|c|c|c}
\hline
\hline
$n$      & $c^{\rm DDH}_n$  & $f^{\rm DDH}_n(r)$ 
         & $c^{\rm EFT}_n$  & $f^{\rm EFT}_n(r)$ & $O^{(n)}_{ij}$ \\
\tableline
1        & $+\frac{g_\pi \, h^1_\pi}{2\sqrt{2} \, m}$ & $f_\pi(r)$  &
$\frac{2\, \mu^2}{\Lambda_\chi^3}\, C_6$  &  
$f_\mu(r)$ & $
({\bm \tau}_i\times {\bm \tau}_j)_z \,({\bm \sigma}_i+{\bm \sigma}_j)\cdot
{\bf X}^{(1)}_{ij,-}$ \\ 
2        & $-\frac{g_\rho \, h^0_\rho}{ m}$  & $f_\rho(r)$  & 0  & 0  & 
${\bm \tau}_i \cdot {\bm \tau}_j \,({\bm \sigma}_i-{\bm \sigma}_j)\cdot
{\bf X}^{(2)}_{ij,+} $          \\
3        & $-\frac{g_\rho \, h^0_\rho(1+\kappa_\rho)}{ m}$ & $f_\rho(r)$  & 0 & 0  & 
${\bm \tau}_i \cdot {\bm \tau}_j \,({\bm \sigma}_i\times{\bm \sigma}_j)\cdot {\bf X}^{(3)}_{ij,-} $      \\
4        & $-\frac{g_\rho \, h^1_\rho}{2\, m}$ & $f_\rho(r)$  & $\frac{ \mu^2}{\Lambda_\chi^3}\, (C_2+C_4)$  &  $f_\mu(r)$ & $
({\bm \tau}_i + {\bm \tau}_j)_z\,({\bm \sigma}_i-{\bm \sigma}_j)\cdot 
{\bf X}^{(4)}_{ij,+} $          \\
5        & $-\frac{g_\rho \, h^1_\rho(1+\kappa_\rho)}{2\, m}$ & $f_\rho(r)$ & 0 & 0 &
$({\bm \tau}_i + {\bm \tau}_j)_z \,({\bm \sigma}_i\times{\bm \sigma}_j)\cdot 
{\bf X}^{(5)}_{ij,-} $   \\
6        & $-\frac{g_\rho \, h^2_\rho}{2 \sqrt{6}\, m}$ & $f_\rho(r)$  & $-\frac{2\, \mu^2}{\Lambda_\chi^3}\, 
C_5$  & $f_\mu(r)$  &
$ (3\, \tau_{i,z} \tau_{j,z}-{\bm \tau}_i \cdot {\bm \tau}_j)
 \,({\bm \sigma}_i-{\bm \sigma}_j)\cdot 
{\bf X}^{(6)}_{ij,+} $       \\
7        & $-\frac{g_\rho \, h^2_\rho(1+\kappa_\rho)}{2\sqrt{6}\, m}$ & $f_\rho(r)$ & 0 & 0 & $ (3\,  \tau_{i,z} \tau_{j,z}-{\bm \tau}_i 
\cdot {\bm \tau}_j) \,({\bm \sigma}_i\times {\bm \sigma}_j)\cdot 
{\bf X}^{(7)}_{ij,-} $  \\
8        & $-\frac{g_\omega \, h^0_\omega}{ m}$  & $f_\omega(r)$  & $\frac{2\, \mu^2}{\Lambda_\chi^3} 
\, C_1$   & $f_\mu(r)$ & $({\bm \sigma}_i-{\bm \sigma}_j)\cdot {\bf X}^{(8)}_{ij,+} $          \\
9        & $-\frac{g_\omega \, h^0_\omega (1+\kappa_\omega)}{ m}$ & $f_\omega(r)$ &  $\frac{2\, \mu^2}{\Lambda_\chi^3} \,\tilde{C}_1$   & $f_\mu(r)$ & $ ({\bm \sigma}_i\times {\bm \sigma}_j)\cdot 
{\bf X}^{(9)}_{ij,-} $   \\
10       & $-\frac{g_\omega \, h^1_\omega}{2\, m}$ & $f_\omega(r)$  &  0 & 0 & $
({\bm \tau}_i + {\bm \tau}_j)_z\, ({\bm \sigma}_i-{\bm \sigma}_j)\cdot 
{\bf X}^{(10)}_{ij,+} $           \\
11       & $-\frac{g_\omega \, h^1_\omega(1+\kappa_\omega)}{2\, m}$ & $f_\omega(r)$  &  0 & 0 & $ ({\bm \tau}_i + {\bm \tau}_j)_z\, ({\bm \sigma}_i\times {\bm \sigma}_j)\cdot 
{\bf X}^{(11)}_{ij,-} $  \\
12       & $-\frac{g_\omega h^1_\omega -g_\rho h^1_\rho}{2\, m}$ & $f_\rho(r)$
& 0 &  0 & 
$ ({\bm \tau}_i -{\bm \tau}_j)_z \,({\bm \sigma}_i+{\bm \sigma}_j)\cdot 
{\bf X}^{(12)}_{ij,+} $           \\
13        & $-\frac{g_\rho \, h^{\prime 1}_\rho}{2\, m}$ & $f_\rho(r)$  & 0 &  0 & $
({\bm \tau}_i\times {\bm \tau}_j)_z \,({\bm \sigma}_i+{\bm \sigma}_j)\cdot {\bf X}^{(13)}_{ij,-}$ \\
\hline
\hline
\end{tabular}
\caption{Components of the DDH and EFT models for the
parity-violating potential.  The vector operators ${\bf X}^{(n)}_{ij,\mp}$ and
functions $f_x(r)$, $x=\pi,\, \rho,\, \omega,\, \mu$, are defined in
Eqs.~(\protect\ref{eq:x+})--(\protect\ref{eq:x-}) and
Eqs.~(\protect\ref{eq:fx})--(\ref{eq:fmu}), respectively.
As outlined in the Appendix~\ref{app:fierz}, only 5 operators and low-energy
constants enter the pionless EFT interaction at the leading order, and in
this paper they have been chosen to correspond to the rows 1, 
4, 6, 8 and 9.
.}
\label{tb:tab1}
\end{table}
\begin{table}[b]
\begin{tabular}{c|ccc|c|c}
\hline
\hline
        & $\Lambda_\pi$ & $\Lambda_\rho$  & $\Lambda_\omega$ & & $\mu$ \\
	\hline
DDH-I   &   1.72   &  1.31     & 1.50      &  EFT-I  & 0.138   \\
DDH-II  & $\infty$ & $\infty$  & $\infty$  &  EFT-II & 1.0    \\  
\hline
\hline
\end{tabular}
\caption{The two different sets of values, in GeV units, for
the short-range cutoffs $\Lambda_x$ (mass $\mu$) used in the
DDH (EFT) model.  Note that the masses $m_\pi$, $m_\rho$, and
$m_\omega$ are taken respectively as 0.138, 0.771, and 0.783
in units of GeV. The row with $\Lambda_x=\infty$ corresponds
to point-like couplings.}
\label{tb:tab2}
\end{table}
\section{Calculation}
\label{sec:cal}

The relevant matrix elements which need to be
calculated in Eq.~(\ref{eq:eq1}) are given by
\begin{equation}
{^{(-)}\!\langle} p;1S;J,J_z\,\mid v^{\rm PV}
\mid p;0J; J,J_z\rangle^{(+)}= 3
\int{\rm d}{\bf x} \, {\rm d}{\bf y}\,
\left[\psi^{\, (-)}_{1S;JJ_z}({\bf x},{\bf y})\right]^\dagger
v^\alpha_{23} \,\psi^{\, (+)}_{0J;JJ_z}({\bf x},{\bf y}) \ ,
\label{eq:eq2}
\end{equation}
where ${\bf x}\equiv {\bf r}_2-{\bf r}_3$ and ${\bf y}\equiv {\bf r}_1-
({\bf r}_2+{\bf r}_3)/2 $ is the $n$-$d$ vector separation defined
earlier.  The factor of 3 on the right-hand-side simply accounts
for the three identical contributions associated with the sum over
pairs included in $v^{\rm PV}$,
\begin{equation}
v^{\rm PV}  =\sum_{i<j} v^\alpha_{ij} \ ,
\end{equation}
where $\alpha$=DDH or EFT.  

The (antisymmetric) wave functions $\psi({\bf x},{\bf y})$,
dropping subscripts and superscripts for brevity, are written as
\begin{equation}\label{eq:decijk}
 \psi({\bf x},{\bf y}) = \sum_{ijk\ {\rm cyclic}}
  \psi_{i;jk}({\bf x}_i,{\bf y}_i) \ ,
\end{equation}
where
\begin{equation}
 {\bf x}_i={\bf r}_j-{\bf r}_k\ ,\qquad 
 {\bf y}_i={\bf r}_i-  ( {\bf r}_j+{\bf r}_k )/2 \ ,
\end{equation}
$(ijk)$=(123), (231), and (312) are the three cyclic permutations of the
particles, and ${\bf x}_1 \equiv{\bf x}$ and ${\bf y}_1\equiv{\bf y}$.
The decomposition given in Eq.~(\ref{eq:decijk}) corresponds to the three
possible partitions of three nucleons 
in 1+2 clusters---indeed, the asymptotic wave functions in
Eq.~(\ref{eq:psia}) have precisely this form.  Note that in the
integral of Eq.~(\ref{eq:eq2}) there are terms like
(again in a schematic notation)
 \begin{equation}
 \int{\rm d}{\bf x} \, {\rm d}{\bf y} \,
 \psi_{1;23}^\dagger({\bf x}_1,{\bf y}_1) \, v_{23}^\alpha\, 
\psi_{1;23}({\bf x}_1,{\bf y}_1) \ ,
 \end{equation}
corresponding to the partition nucleon 1 and nucleons 23 in the
initial and final $n$-$d$ states, and with the PV
interaction acting on nucleons 23.  These $n$-$d$ continuum
wave functions have ``core'' parts which vanish at $\infty$, and
asymptotic parts oscillating in the $y_1$ variable.
The integrals involving core parts of either the left
or right (or both) wave functions are converging, while
the integral involving the left and right asymptotic wave
functions vanishes, since the integrand is odd under
inversion ${\bf y}_1 \rightarrow -{\bf y}_1$.  The
remaining partitions give converging contributions to the
integral, since the deuteron cluster in the left and right
wave functions is made up of different nucleons and
$\psi_{2;31}({\bf x}_2,{\bf y}_2), \psi_{3;12}({\bf x}_3,{\bf
  y}_3)\rightarrow 0$ as ${\bf y}_1 \rightarrow\infty$.

\begin{figure}[t]
\includegraphics[width=6in]{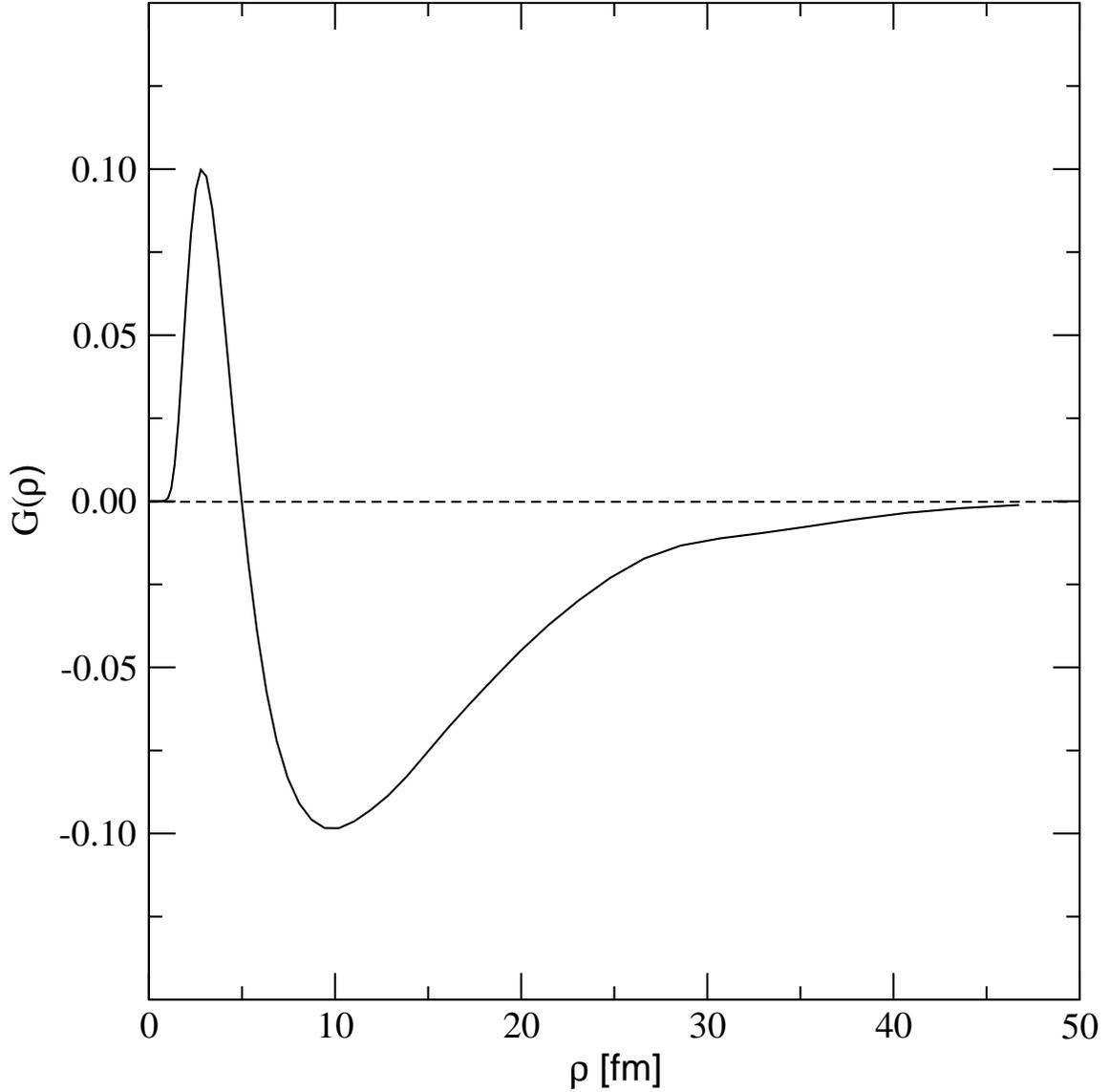}
\caption{The function $G(\rho)$, defined so that 
$\int^\infty_0 {\rm d}\rho\, G(\rho)$ is the matrix element of the operator $O^{(1)}$
between the $^2$S$_{1/2}$ and $^2$P$_{1/2}$ $n$-$d$ states at zero
energy, plotted as function of the hyper-radius
$\rho = \sqrt{x^2+4\, y^2/3}$.} 
\label{fig:fig1}
\end{figure}

The wave functions for an assigned spatial
configuration $({\bf x},{\bf y})$ are expanded
on a basis of $8\times 3$ spin-isospin states for the three nucleons as
\begin{equation}
\psi({\bf x},{\bf y}) = \sum_{a=1}^{24} \psi_a({\bf x},{\bf y}) \mid\! a\rangle \ ,
\end{equation}
where the components $\psi_a({\bf x},{\bf y})$ are generally complex
functions, and the basis states $\mid\! a\rangle$=
$\mid\! (n\! \downarrow)_1 (p\! \downarrow)_2 (n\! \downarrow)_3\rangle$,
$\mid\! (n\! \downarrow)_1 (n\! \downarrow)_2 (p\! \downarrow)_3\rangle$, and
so on.  Matrix elements of the PV potential components are written
schematically as
\begin{equation}
\langle f\!\mid O \mid \! i \rangle=\sum_{a,b=1}^{24} \int 
{\rm d}{\bf x} \, {\rm d}{\bf y}
\, \psi^*_{f,a}({\bf x},{\bf y}) \left[ O({\bf x})\right]_{ab} 
\psi_{i,b}({\bf x},{\bf y}) \ ,
\label{eq:mci}
\end{equation}
where $\left[ O({\bf x})\right]_{ab}$ denotes the matrix representing in
configuration space any of the components in Table~\ref{tb:tab1}.  Note
that the operators ${\bf X}^{(n)}_{23,\mp}$ occurring in $v^\alpha_{23}$
are conveniently expressed as
\begin{equation}\label{eq:xcomm}
  {\bf X}_{23,-}^{(n)}=\hat{\bf x} \, f^\prime_n(x) \ , \qquad
  {\bf X}_{23,+}^{(n)} = -{\rm i} \left[ 2\, f_n(x) \,{\bm \nabla}_x +
\hat{\bf x} \, f^\prime_n(x) \right] \ ,
\end{equation}
where the gradient operator ${\bm \nabla}_x= ({\bm \nabla}_2-{\bm \nabla}_3)/2$
acts on the right (initial) wave function, and $f^\prime(x) ={\rm d}f(x)/{\rm
d}x$.  Gradients are discretized as 
\begin{equation}
\nabla_{i,\alpha} \psi ({\bf x},{\bf y})\simeq 
\frac{\psi(\dots {\bf r}_{i}+\delta\, \hat{\bf e}_\alpha \dots)
-\psi(\dots {\bf r}_{i}-\delta\, \hat{\bf e}_\alpha \dots)}{2\,\delta} \ ,
\end{equation}
where $\delta$ is a small increment and $\hat{\bf e}_\alpha$ is a unit
vector in the $\alpha$-direction.  Matrix multiplications in the
spin-isospin space are performed exactly with the techniques developed
in Ref.~\cite{Schiavilla89}.  The problem is then reduced to the evaluation
of the spatial integrals, which is efficiently carried out by a combination of
Monte Carlo (MC) and standard quadratures techniques.  We write
\begin{equation}
 \langle f\!\mid O \mid \! i \rangle = \int{\rm d} \hat{\bf x}
\,{\rm d} \hat{\bf y} \, F(\hat{\bf x},\hat{\bf y}) 
 \simeq \frac{1}{N_c} \sum_{c=1}^{N_c} F(c) \ ,
\end{equation}
where the $c$'s denote uniformly sampled directions
$(\hat{\bf x},\hat{\bf y})$ (total number $N_c$).  For each such
configuration $c$, the function $F$ is obtained by Gaussian
integrations over the $x$ and $y$ variables, {\it i.e.}
\begin{equation}
F(c)= (4 \pi)^2 \sum_{a,b=1}^{24} \int_0^\infty {\rm d}x \, x^2
\int_0^\infty {\rm d}y\, y^2\, \psi^*_{f,a}({\bf x},{\bf y})
\left[ O({\bf x})\right]_{ab}  
\psi_{i,b}({\bf x},{\bf y}) \ .
\end{equation}
Convergence in the $x$ and $y$ integrations requires of the
order of 40 Gaussian points, distributed over a non-uniform grid
extending beyond 40 fm, while $N_c$ of the order of a few thousands 
is sufficient to reduce the statistical error in the MC integration
at the percent level.  Thus, the present method turns
out to be computationally intensive.  In this respect, we note
that a direct MC evaluation of the six-dimensional integral in
Eq.~(\ref{eq:mci}) by Metropolis sampling from a probability density
function $W({\bf x},{\bf y}) \propto \, \mid \! \psi_0({\bf x},{\bf y})\!\mid^2$,
where $\psi_0$ is the triton bound-state wave
function, turned out to be impractical.  This was also the case
for other choices of $W$, such as $W({\bf x},{\bf y}) \propto
x^\lambda y^\nu \mid\! \phi({\bf x})\!\mid^2  \mid\! \phi({\bf y})\!\mid^2 $,
where $\phi$ is the deuteron wave function.  The reason for this
difficulty can be appreciated by examining Fig.~\ref{fig:fig1},
in which the integrand associated with the pion-range component
of the DDH potential, specifically the matrix element of $O^{(1)}$
between the $^2$S$_{1/2}$ and $^2$P$_{1/2}$ $n$-$d$ states at zero
energy, is plotted as function of the hyper-radius $\rho =\sqrt{x^2+4\, y^2/3}$.
Note the change of sign in this function, due to the node in the
continuum $^2$S$_{1/2}$ state (thus ensuring its orthogonality to the
$^3$H bound state), and its long-range character.  This last feature,
in particular, makes it difficult for an efficient sampling of the large $\rho$
region with the importance functions $W$ mentioned above.

The computer codes were successfully tested by carrying out
a calculation based on Gaussian wave functions for the initial
and final states, as described in the following subsection.

\subsection{Test calculation} 
\label{sec:calt}

We have performed calculations with Gaussian-type
wave functions, for which the matrix elements can be computed
analytically.  In Eq.~(\ref{eq:eq2}), we have replaced the
realistic $n$-$d$ wave functions with the following ones
\begin{equation}
 \widetilde{\psi}_{LS;JJ_z}({\bf x},{\bf y}) =
\frac{ {\rm exp}(-\beta \, \rho^{\, 2})}{\sqrt{4\pi}} \Omega_{LS;JJ_z} \ ,
\label{eq:psit1}
\end{equation}
with
\begin{equation}
\Omega_{LS;JJ_z}= \sum_{ijk\, {\rm cyclic}} y_i^L \, F_{LSJ}(y_i) \, \Big [ Y_{L}(\hat{\bf y}_i)
  \, \otimes \big[s_i S_{jk}\big]_S\Big ]_{JJ_z}\;
   \big[t_i T_{jk}\big]_{TT_z}  \ ,
  \label{eq:psit2} 
\end{equation}
where $\rho$ is the hyper-radius (independent of the particle permutation
considered), $s_i$ ($t_i$) denotes the spin (isospin) state of particle $i$, and
$S_{jk}$ ($T_{jk}$) the spin (isospin) state of particles $jk$. 
We have considered only states of total isospin  $T,T_z$=1/2,\hbox{--1/2}, appropriate
to describe the $n$-$d$ channel---possible admixtures of $T$=3/2 components induced
by isospin symmetry breaking terms in the strong and electromagnetic interactions
are expected to give negligible contributions to PV observables.

The quantum numbers $S_{jk}$ and $T_{jk}$ are respectively taken to be 1 and 0,
while the factors $F_{LSJ}(y_i)$ selected for the various channels are
reported in Table~\ref{tb:tab3}.  Note that $F_{LSJ}(y_i)$=1 except
for $L$=0 and $S$=$J$=3/2, since in this case the spin part
$\big[s_i S_{jk}\big]_{3/2}$ is totally symmetric, and it is impossible
to construct a fully antisymmetric wave function with only the isospin
part.  Hence, the choice $F(y_i)$=$y_i^{\, 2}$.  In all other channels,
non-vanishing functions $\Omega_{LS;JJ_z}$ are obtained by simply
taking $F_{LSJ}(y_i)$=1. 

\begin{table}[b]
\begin{tabular}{lll c}
\hline
\hline
$L\ \ $ & $S\ \ $ & $J\ \ $ & $F_{LSJ}(y_i)$ \\
\hline
 $0$ & ${1\over2}$ & ${1\over 2}$ & $1$ \\
 $0$ & ${3\over2}$ & ${3\over 2}$ & $y_i^{\, 2}$ \\
 $1$ & ${1\over2}$ & ${1\over 2}$ & $1$ \\
 $1$ & ${3\over2}$ & ${1\over 2}$ & $1$ \\
 $1$ & ${1\over2}$ & ${3\over 2}$ & $1$ \\
 $1$ & ${3\over2}$ & ${3\over 2}$ & $1$ \\
\hline
\hline
\end{tabular}
\caption{Quantum numbers and factor $F_{LSJ}(y_i)$ entering the Gaussian wave
functions given in Eq.~(\ref{eq:psit1}).}
\label{tb:tab3}
\end{table}
The analytical calculation is simplified considerably by expressing
the function $\Omega_{LS;JJ_z}$, which involves the three set of
coordinates $({\bf x}_1,{\bf y}_1)$, $({\bf x}_2,{\bf y}_2)$, and 
$({\bf x}_3,{\bf y}_3)$ corresponding to the three partitions
(1,23), (2,31), and (3,12), in terms of linear combinations
of functions of, say, the first set of spatial variables, {\it i.e.}
${\bf x}\equiv {\bf x}_1$ and ${\bf y}\equiv {\bf y}_1$ as in
Eq.~(\ref{eq:eq2}), and spin-isospin states of pair 23 only.  We find:
\begin{equation}
 \Omega_{LS;JJ_z} = \sum_{n_x,n_y,\ell_x,S_{x},j_x,\ell_y,j_y,T_{x}}
  c^{LSJ}_{n_x,n_y,\ell_x,S_{x},j_x,\ell_y,j_y,T_{x}}\;
   \Omega^{JJ_z;TT_z}_{n_x,n_y,\ell_x,S_{x},j_x,\ell_y,j_y,T_{x}}\ ,
\end{equation}
where 
\begin{equation}
 \Omega^{JJ_z;TT_z}_{n_x,n_y,\ell_x,S_{x},j_x,\ell_y,j_y,T_{x}}=
  x^{\ell_x+n_x} y^{\ell_y+n_y} 
 \Big[ \big[Y_{\ell_x}(\hat{\bf x}) S_{x}\big]_{j_x}
  \, \otimes 
  \big[Y_{\ell_y}(\hat{\bf y}) s_1 \big]_{j_y}
  \Big]_{JJ_z}\;
   \big[T_{x} t_1\big]_{TT_z}  \ ,
  \label{eq:psit4} 
\end{equation}
and $n_x$ are $n_y$ are non-negative integers, $S_x$ and $T_x$ are
the spin and isospin states of pair 23, and
$c^{LSJ}_{n_x,n_y,\ell_x,S_{x},j_x,\ell_y,j_y,T_{x}}$ are numerical
factors, which can easily be expressed
in terms of Wigner coefficients.

It is now relatively simple to compute matrix elements of the
operators $O_{23}^{(n)}$ in Table~\ref{tb:tab1} between states 
$\Omega^{JJ_z;TT_z}_{n_x,n_y,\ell_x,S_{x},j_x,\ell_y,j_y,T_{x}}$
by making use of the following relations
\begin{equation}
\int {\rm d}\hat{\bf x}\, 
   \big[Y_{\ell_x}(\hat{\bf x}) S_{x}\big]^\dag _{j_x m_x} \,
    ({\bm \sigma}_2 + {\bm \sigma}_3)\cdot \hat{\bf x} \, 
   \big[Y_{\ell_x^\prime}(\hat{\bf x}) S_{x}^\prime\big]_{j_x m_x} =
   -2\sqrt{\frac{j_x+1/2 \pm 1/2}{2j_x+1}}\, 
   \delta_{\ell_x,\, j_x}\, \delta_{S_x,1}\, \delta_{S^\prime_x,1}
   \, \delta_{\ell_x^\prime ,\,j_x\mp 1}\ ,
\end{equation}
and similar ones for the spin-space and isospin operators
occurring in the potential components $O_{23}^{(n)}$.  In particular,
those for $({\bm \sigma}_2 - {\bm \sigma}_3)\cdot{\bm \nabla}_x $ and 
$({\bm \sigma}_2 \times {\bm \sigma}_3)\cdot \hat{\bf x}$ are listed
in Sec.~III.E of Ref.~\cite{Carlson02}.  The remaining task of
integrating over the radial variables $x$ and $y$ is carried out
accurately by means of Gauss quadrature formulae.

The results of the ``analytical'' calculation just described are
found to be in agreement, for each of the 13 components of the PV
potential, with those produced by a MC calculation using the same
input wave functions as in Eq.~(\ref{eq:psit1}).  The spin-isospin
algebra is performed as outlined in the previous section; however,
because of the short-range character of the wave functions in this
case, spatial configurations turn out to be efficiently sampled with
the Metropolis algorithm by a probability density proportional to
${\rm exp}(-2\, \beta \rho^2\, )$. In a typical calculation, the values 
$\beta=0.10$--$0.25$ fm${}^{-2}$ have been selected.

\section{Results and Conclusions}
\label{sec:res}

The results for the coefficients $I^\alpha_n$ in Eq.~(\ref{eq:in}),
obtained with $n$-$d$ continuum wave functions corresponding to
the AV18 and AV18/UIX strong-interaction Hamiltonians, are reported
for the DDH and pionless EFT PV potentials in Tables~\ref{tb:tab4}
and~\ref{tb:tab5}, respectively.  The labels I and II for the columns
in each of these tables refer to the two sets of cutoff parameters, as
given in Table~\ref{tb:tab2}.  The subscript $n$ in $I^\alpha_n$
specifies the operators as listed in Table~\ref{tb:tab1}.  Note that
results for $n$=13 are not reported, since the coupling constant
$h_\rho^{1\, \prime}$ multiplying these contributions has been estimated
to be considerably smaller~\cite{Holstein81} than the ``best'' values
estimated for the other coupling constants in the original DDH
reference~\cite{Desplanques80}.
\begin{table}[bth]
\begin{tabular}{c|d|d||d|d}
  & \multicolumn{2}{c}{DDH-I} & \multicolumn{2}{c}{DDH-II} \\
\hline
\hline
$n$ & {\rm AV18} & {\rm AV18/UIX} & {\rm AV18} & {\rm AV18/UIX}  \\
\tableline
1 &  $ 0.256E+03$  & $ 0.270E+03$  &  $ 0.257E+03$  & $ 0.274E+03$  \\
2 &  $--0.444E+01$  & $--0.691E+01$  &  $--0.719E+01$  & $--0.118E+02$  \\
3 &  $ 0.444E+01$  & $ 0.401E+01$  &  $ 0.732E+01$  & $ 0.761E+01$  \\
4 &  $--0.231E+01$  & $--0.881E+00$  &  $--0.332E+01$  & $--0.148E+01$ \\
5 &  $--0.247E+01$  & $--0.122E+01$  &  $--0.387E+01$  & $--0.234E+01$ \\
8 &  $ 0.420E+01$  & $ 0.362E+01$  &  $ 0.543E+01$  & $ 0.516E+01$ \\
9 &  $ 0.111E+01$  & $--0.117E+00$  &  $ 0.136E+01$  & $--0.189E+00$ \\
10&  $--0.253E+01$  & $--0.991E+00$  &  $--0.314E+01$  & $--0.141E+01$ \\
11&  $--0.280E+01$  & $--0.144E+01$  &  $--0.369E+01$  & $--0.225E+01$ \\
12&  $ 0.316E+01$  & $ 0.339E+01$  &  $ 0.455E+01$  & $ 0.546E+01$ \\
\hline
\hline
\end{tabular}
\caption{The coefficients $I^{\rm DDH}_n$ (in fm) in Eq.~(\protect\ref{eq:in})
corresponding to the DDH PV potential in combination with the AV18 and
AV18/UIX strong-interaction potentials for the two different sets of
cutoff parameters given in Table~\protect\ref{tb:tab2}.  The statistical
errors associated with the Monte Carlo integrations are not shown, but
are typically at the 1-2\% level.  Note that the coefficients
$I^{\rm DDH}_{n=6,7}$ vanish because of isospin selection rules.}
\label{tb:tab4}
\end{table}

A quick glance at Table~\ref{tb:tab4} makes it clear that the
matrix element of the long-range component of the DDH potential
due to pion exchange is almost two orders of magnitude larger
than that of its short-range components induced by vector-meson
exchanges.  It is also clear that this contribution is fairly
insensitive to the choice of cutoffs as well as of input strong-interaction
Hamiltonians used to generate the $n$-$d$ S- and P-wave channel
states.  In particular, one should note that the AV18 and AV18/UIX
models predict rather different values for the scattering length
in the $^2$S$_{1/2}$ channel~\cite{Kievsky08}.  On the other hand, predictions for
the $^4$S$_{1/2}$ scattering length are very close to each other.
This is reflected in the contributions to $I_1^{\rm DDH}$ due
to the doublet and quartet S-wave channels.  We find that for the
case of DDH-I, for example, these contributions for the AV18 and
AV18/UIX are, respectively, 45.0 fm and 74.2 fm in the $^2$S$_{1/2}$ channel,
and 210.6 fm and 196.6 fm in the $^4$S$_{1/2}$ channel, which add up (up to
truncation errors) to the values 256 fm and 270 fm, reported in the first
row of Table~\ref{tb:tab4}.  We observe that the doublet scattering length
calculated with the AV18/UIX is very close to the experimental
determination of this quantity, while the quartet scattering length, which
is much less sensitive to the presence of three-nucleon interactions,
is found to be in agreement with the experimental value with
both the AV18 and AV18/UIX models.  Hence, it is reasonable to
expect that any strong-interaction Hamiltonian that reproduces
the experimental scattering lengths should lead to predictions
for the long-range component of the DDH potential, which are very
similar to those reported here for the AV18/UIX model.

There is considerable model dependence in the results obtained
for the individual contributions due to vector-meson exchanges.
However, this model dependence has little impact on the neutron
spin rotation, as it can be surmised from Table~\ref{tb:tab6},
where we list predictions obtained for this observable.  We adopt
two different sets of strong- and weak-interaction coupling
constants, namely those reported in Tables~I and~II of Ref.~\cite{Schiavilla04}.
The set of coupling constants denoted as DDH-best (from Table II
of Ref.~\cite{Schiavilla04}) is the ``best value'' set as given
in the original work~\cite{Desplanques80}.  Recently, the $\rho$- and
$\omega$-meson weak-coupling constants were adjusted in Ref.~\cite{Carlson02}
to reproduce precise measurements of the longitudinal asymmetry in
$\vec{p}-p$ elastic scattering (see Table I 
of Ref.~\cite{Schiavilla04}).  The resulting set of coupling constants
is denoted as DDH-adj.  Note that both sets use 
$h^1_\pi$=$4.56\times 10^{-7}$.  Assuming a liquid deuterium density of
$\rho=0.4\times10^{23}$ atoms cm${}^{-3}$, we obtain
${\rm d}\phi/{\rm d}z\approx 0.53\times10^{-7}$ rad-cm${}^{-1}$
($0.56\times10^{-7}$ rad-cm${}^{-1}$) with the
AV18/UIX+DDH-best (AV18/UIX+DDH-adj) model. About 90\% of this prediction
comes from the long-range component of DDH, while its short-range components
contribute the remaining 10\%.  In particular, the reason for the larger
short-range contribution obtained with DDH-adj relative to DDH-best
lies in the fact that this contribution is dominated by the
$n=3$ term in Eq.~(\ref{eq:in}), for which the combination
of coupling constants is  $c_3=-g_\rho\, h^0_\rho\,(1+\kappa_\rho)/m$
(see Table~\ref{tb:tab1}).  In DDH-adj $c_3$ is roughly a factor
of two larger than in DDH-best because of different values for $h^0_\rho$
and the tensor coupling constant $\kappa_\rho$.

It is interesting to note that the neutron spin rotation in
$\vec{n}$-$p$ scattering is predicted to be about one order of
magnitude smaller than in the present case~\cite{Schiavilla04},
which should make its measurement in $\vec{n}$-$d$ scattering 
considerably easier, at least in principle.

\begin{table}[bth]
\begin{tabular}{c|d|d||d|d}
  & \multicolumn{2}{c}{EFT-I} & \multicolumn{2}{c}{EFT-II} \\
\hline
\hline
$n$ & ${\rm AV18}$ & ${\rm AV18/UIX}$ & ${\rm AV18}$ & ${\rm AV18/UIX}$ \\
\tableline
1 &  $ 0.257E+03$  & $ 0.274E+03$ &  $ 0.392E+01$  & $ 0.706E+01$ \\
4 &  $--0.154E+03$  & $--0.508E+02$ &  $--0.127E+01$  & $--0.682E+00$ \\
8 &  $ 0.260E+03$  & $ 0.189E+03$ &  $ 0.233E+01$  & $ 0.251E+01$ \\
9 &  $ 0.244E+00$  & $--0.359E+02$ &  $ 0.557E+00$  & $--0.955E-01$ \\
\hline
\hline
\end{tabular}
\caption{Same as in Table~\protect\ref{tb:tab4} but for the pionless EFT
PV potential.  Note that there are no potential components
with $O^{(n)}_{ij}$ with $n$=2, 3, 5, 7, 10 and 11, and that the
coefficient $I^{\rm EFT}_{n=6}$ vanishes
because of isospin selection rules.}
\label{tb:tab5}
\end{table}

The coefficients $I^{\rm EFT}_n$ for the relevant operators
entering the pionless EFT PV potential, that is $n$=1, 4, 6, 8,
and 9, are reported in Table~\ref{tb:tab5}.  Note that the $n$=6
operator is isotensor, and therefore its matrix element vanishes
here.  The $I^{\rm EFT}_n$ coefficients are all of the same order
of magnitude, since the radial functions are taken to be the same
for all $n$, $f^{\rm EFT}_n(r)=f_\mu(r)$.  Of course, they depend
significantly on the value of the mass $\mu$---either $\mu=m_\pi$
(EFT-I), as appropriate in the present pionless EFT formulation, 
or $\mu=1$ GeV (EFT-II), the scale of chiral symmetry breaking,
as appropriate in the formulation in which pion degrees of freedom
are explicitly retained.  Indeed, in this latter formulation
the leading order component of $v^{\rm PV}$ has the same form as
the pion-exchange term in DDH.

We note that the coefficient $I^{\rm EFT}_{n=1}$ is
weakly dependent on the choice of input strong-interaction Hamiltonian---AV18
or AV18/UIX---in the case of the EFT-I model.  On the other
hand, this is not the case the for $I^{\rm EFT}_{n=4,8,9}$
coefficients. 

Finally, rough estimates have been made for the range of values
allowed for the low-energy constants $C_1$, $C_2+C_4$, $C_5$,
$\tilde C_1$, and $C_6$ in Ref.~\cite{Zhu05}.  However, at the
present time a systematic program for their determination is yet
to be carried out.  In view of this, we refrain here from making EFT-based
predictions for the spin rotation observable.
\begin{table}[bth]
\begin{tabular}{c|d|d||d|d}
  & \multicolumn{2}{c}{AV18} & \multicolumn{2}{c}{AV18/UIX} \\
\hline
\hline
     & ${\rm DDH-best}$  & ${\rm DDH-adj}$ & ${\rm DDH-best}$ & ${\rm DDH-adj}$ \\
\tableline
$\pi$ & 0.457 & 0.457 & 0.485 & 0.485 \\
$\rho$-$\omega$    & 0.063 & 0.092 & 0.046 & 0.075 \\
TOT   & 0.520 & 0.550 & 0.531 & 0.560 \\
\hline
\hline
\end{tabular}
\caption{Spin rotation for $\vec{n}-d$ scattering in units of $10^{-7}$
rad-cm${}^{-1}$ at zero energy, obtained for the DDH-I model assuming a
liquid deuterium density of $\rho=0.4\times10^{23}$ atoms cm${}^{-3}$.
The columns labeled DDH-best (DDH-adj) list the spin rotation predicted
using the ``best'' (``adjusted'') values for the weak-interaction coupling
constants, as reported in Table II (I) of Ref.~\protect\cite{Schiavilla04}.
The row labeled ``$\pi$'' (``$\rho$-$\omega$'') lists the results obtained
by including only the pion ($\rho$ and $\omega$ mesons), while that labeled
``TOT'' gives the total contributions.}
\label{tb:tab6}
\end{table}

In conclusion, we have calculated the neutron spin rotation, induced
by PV components in the nucleon-nucleon potential, in $\vec{n}$-$d$
scattering at zero energy.  We find that this observable is dominated
by the contribution of the long-range part of the PV potential
due to pion exchange, and that the model-dependence of
this contribution is weak.  The predicted ${\rm d}\phi/{\rm d}z$ is
$\simeq 0.5 \times 10^{-7}$ rad-cm$^{-1}$---an order of magnitude
larger than expected in $\vec{n}$-$p$ scattering~\cite{Schiavilla04}.
Thus a measurement of the spin rotation in $\vec{n}$-$d$ scattering
could provide a further constraint, complementary to that coming from
measurements of the photon asymmetry in $\vec{n}$-$p$ radiative capture,
on the strength of the long-range component of the hadronic weak interaction.

\section*{Acknowledgments}

We would like to thank D.\ Markoff for her interest
in the present work.  One of the authors (R.S.) would
also like to thank the Physics Department of the University
of Pisa for the support and warm hospitality extended to
him on several occasions.

The work of R.S.\ is supported by the U.S.\ Department of Energy
under contract DE-AC05-06R23177.  The calculations were made
possible by grants of computing time from the National Energy
Research Supercomputer Center.

\appendix
\section{}
\label{app:fierz}

The leading order PV contact Lagrangian has been written in
Ref.~\cite{Zhu05} as consisting of twelve operators,
\begin{equation}
  {\cal L}= \sum_{i=1}^6 (C_i O_i + \tilde C_i \tilde O_i)\ ,
\end{equation}
where 
\begin{equation}
\begin{array}{ll}
 O_1 =  \bar \psi  1 \gamma^\mu \psi \bar \psi  1 \gamma_\mu
\gamma_5 \psi\ , &  \tilde O_1 =  \bar \psi 1
 \gamma^\mu \gamma_5 \psi \partial^\nu ( \bar \psi 1 \sigma_{\mu\nu}
\psi) \ ,  \\
 O_2 =  \bar \psi  1 \gamma^\mu \psi \bar \psi \tau_3 \gamma_\mu
\gamma_5 \psi\ ,   & \tilde O_2 =  \bar \psi 1
 \gamma^\mu \gamma_5 \psi \partial^\nu ( \bar \psi \tau_3 \sigma_{\mu\nu}
\psi)\ ,  \\
 O_3 =  \bar \psi  \tau^a \gamma^\mu \psi \bar \psi  \tau^a \gamma_\mu
\gamma_5 \psi\ ,  & \tilde O_3 =  \bar \psi \tau^a
 \gamma^\mu \gamma_5 \psi \partial^\nu ( \bar \psi \tau^a \sigma_{\mu\nu}
\psi)\ ,  \\
 O_4 =  \bar \psi  \tau_3 \gamma^\mu \psi \bar \psi 1 \gamma_\mu
\gamma_5 \psi\ ,  & \tilde O_4 =  \bar \psi \tau_3
 \gamma^\mu \gamma_5 \psi \partial^\nu ( \bar \psi 1 \sigma_{\mu\nu}
\psi)\ ,   \\
 O_5 =  {\cal I}_{ab} \bar \psi  \tau^a \gamma^\mu \psi \bar \psi  \tau^b \gamma_\mu
\gamma_5 \psi\ , & \tilde O_5 = {\cal I}_{ab} \bar \psi \tau^a
 \gamma^\mu \gamma_5 \psi \partial^\nu ( \bar \psi \tau^b \sigma_{\mu\nu}
\psi)\ ,   \\
 O_6 =  i \epsilon^{ab3} \bar \psi  \tau^a  \psi \bar \psi  \tau^b \gamma_5
 \psi\ ,  & \tilde O_6 = i \epsilon^{ab3}  \bar \psi \tau^a
\gamma^\mu  \psi \partial^\nu ( \bar \psi \tau^b \sigma_{\mu\nu} \gamma_5
\psi)\ ,
\end{array}
\end{equation}
where the nucleon field $\psi$ is a doublet in isospin space, and the operators
$1$ and $\tau^a$ above act on this space.  In addition, 
${\cal I}_{ab}$ is a $3\times3$ diagonal matrix with elements $(1,1,-2)$.
Using Fierz transformations and
fields' equations of motion, the following relations may be shown to
hold \cite{Girlanda08}, 
\begin{equation}
\begin{array}{l}
O_3=O_1\ ,\\
O_2 - O_4=2 O_6\ ,\\
\tilde O_3 = 2 m ( O_1 + O_3) - 3 \tilde O_1\ ,\\
\tilde O_2 + \tilde O_4 = m ( O_2 + O_4)\ , \\
\tilde O_2 - \tilde O_4 = -2 m O_6 - \tilde O_6\ ,\\
\tilde O_5 = O_5\ ,
\end{array}
\end{equation}
so that the number of independent operators can be reduced to
six.  Moreover, since the
relativistic operators $O_6$ and $\tilde O_6$ give rise to the same
structure in the leading order of the non-relativistic expansion, as
already observed in Ref.~\cite{Zhu05}, the PV two-nucleon
non-relativistic contact Lagrangian consists of only five operators,
\begin{eqnarray}
{\cal L}_{PV,NN} &=& \frac{1}{\Lambda_\chi^3} \left\{ C_1 ( N^\dagger
\vec{\sigma} N \cdot N^\dagger i \stackrel{\leftrightarrow}{\nabla} N
- N^\dagger
 N  N^\dagger i \stackrel{\leftrightarrow}{\nabla} \cdot \vec{\sigma} N
 ) \nonumber \right.
 - \tilde C_1\, \epsilon_{ijk}\, N^\dagger\, \sigma^i\, N \nabla^j
( N^\dagger \, \sigma^k \, N) \nonumber \\
&&- (C_{2}+C_4) \epsilon_{ijk}\, [N^\dagger \,\tau_3\, \sigma^i N \nabla^j
( N^\dagger\,\sigma^k  \,N) + N^\dagger \, \sigma^i\, N \nabla^j
( N^\dagger\, \tau_3\, \sigma^k\,  N)] \nonumber \\
&& 
- \tilde C_5 {\cal I}_{ab}\, \epsilon_{ijk}\, N^\dagger \tau^a \sigma^i N \nabla^j
\, ( N^\dagger \tau^b \sigma^k  N) \left.
+C_6\, \epsilon^{ab3}\, \vec{\nabla} (N^\dagger \tau^a N )\cdot  N^\dagger \tau^b
\vec{\sigma} N  \right\},
\end{eqnarray}
where the notations of Ref.~\cite{Zhu05} have been chosen for the
coupling constants.
Accordingly, the pionless EFT potential in coordinate space reads as
\begin{eqnarray}\label{3}\nonumber
v^{\rm EFT}_{ij} &=&
{2\mu^2 \over \Lambda_\chi^3} \left\{
\left[ C_1 + (C_2+C_4) \left({\tau_i +\tau_j \over 2}\right)_z
+C_5\, {\cal I}_{ab} \tau_i^a \tau_j^b\right]
 \left( {\bm \sigma}_i -{\bm \sigma}_j\right)\cdot \left[{\bf p}_{ij}\, ,\, f_\mu(r)\right]_+
\right.\\
\nonumber &&
\left.
\qquad +
i\, {\tilde C}_1 \left( {\bm \sigma}_i\times {\bm \sigma}_j \right) \cdot
\left[ {\bf p}_{ij}\, ,\, f_\mu(r)\right]_-
 +i\, C_6 \epsilon^{ab3} \tau_i^a \tau_j^b
\left( {\bm \sigma}_i +{\bm \sigma}_j\right)\cdot
\left[{\bf p}_{ij}\, ,\, f_\mu(r)\right]_- \right\} \ ,
\label{eq:vpvsr}
\end{eqnarray}
where ${\bf p}_{ij}$ is the relative momentum operator and $f_\mu(r)$, defined
in Eq.~(\ref{eq:fmu}), is introduced to regularize the potential at short-range.

\newpage


\begin{thebibliography}{100}
%
\bibitem{Carlson02}
J.\ Carlson, R.\ Schiavilla, V.R.\ Brown, and B.F.\ Gibson,
Phys.\ Rev.\ C {\bf 65}, 035502 (2002).
%
\bibitem{Schiavilla04}
R.\ Schiavilla, J.\ Carlson, and M.\ Paris,
Phys.\ Rev.\ C {\bf 70}, 044007 (2004).
%
\bibitem{Desplanques80}
B.\ Desplanques, J.F.\ Donoghue, B.R.\ Holstein,
Ann.\ Phys.\ (N.Y.) 124, 449 (1980).
%
\bibitem{Ramsey06}
M.J.\ Ramsey-Musolf and S.A.\ Page,
Ann.\ Rev.\ Nucl.\ Part.\ Sci.\ {\bf 56}, 1 (2006).
%
\bibitem{Zhu05}
S.-L.\ Zhu, C.M.\ Maekawa, B.R.\ Holstein, M.J.\ Ramsey-Musolf, and U.\ van Kolck,
Nucl.\ Phys.\ {\bf A748}, 435 (2005).
%
\bibitem{Kievsky94} A. Kievsky, S. Rosati, and M. Viviani,
Nucl. Phys. {\bf A577},  511 (1994)
%
\bibitem{Kievsky04}
A. Kievsky, M. Viviani, and L.E. Marcucci,
Phys. Rev. {\bf C69}, 014002 (2004)

\bibitem{Kievsky08} A. Kievsky, S. Rosati, M. Viviani, L.E. Marcucci, and
  L. Girlanda, J. Phys. G {\bf 35}, 063101 (2008)
%
\bibitem{Wiringa95}
R.B.\ Wiringa, V.G.J.\ Stoks, and R.\ Schiavilla,
Phys.\ Rev.\ C {\bf 51}, 38 (1995).
%
\bibitem{Pudliner95}
B.S.\ Pudliner {\it et al.},
Phys.\ Rev.\ Lett.\ {\bf 74}, 4396 (1995).
%
\bibitem{Carlson98}
J.\ Carlson and R.\ Schiavilla,
Rev.\ Mod.\ Phys.\ {\bf 70}, 743 (1998).
%
\bibitem{Marcucci06} 
L.E.\ Marcucci, K.M.\ Nollett, R.\ Schiavilla, and R.B.\ Wiringa,
Nucl.\ Phys.\ {\bf A777}, 111 (2006).
%
\bibitem{Fermi50}
E.\ Fermi,
{\it Nuclear Physics} (The University of Chicago Press, Chicago, 1950).
%
\bibitem{Viviani96}
M.\ Viviani, R.\ Schiavilla, and A.\ Kievsky,
Phys.\ Rev.\ C {\bf 54}, 534 (1996).
%
\bibitem{Holstein81}
B.R.\ Holstein,
Phys.\ Rev.\ D {\bf 23}, 1618 (1981).
%
\bibitem{Girlanda08}
L.\ Girlanda,
arXiv:0804.0772.
%
\bibitem{Schiavilla89}
R.\ Schiavilla, V.R.\ Pandharipande, and D.O.\ Riska,
Phys.\ Rev.\ C {\bf 40}, 2294 (1989).
%
\end{thebibliography}
\end{document}